\documentclass[usenatbib]{mn2e}
\usepackage{graphicx}
\usepackage{pstricks}
\date{Accepted}
\pagerange{00--00} \pubyear{2006}
\newlength{\figurewidth}
\newcommand{\revised}[1]{{#1}}
\title[An abnormal massive-star-ONC-IMF?]{A highly abnormal massive star mass function in the Orion Nebula
cluster and the dynamical decay of trapezia systems}
\author[J.~Pflamm-Altenburg, P.~Kroupa]
{J.~Pflamm-Altenburg$^1$$^2$\thanks{email: jpflamm@astro.uni-bonn.de,
    pavel@astro.uni-bonn.de} and 
  P.~Kroupa$^1$$^2$\footnotemark[1]\\
  $^1$ Argelander-Institut f\"ur Astronomie\thanks{Founded by 
    merging of the Institut f\"ur Astrophysik
    und Extraterrestrische Forschung, the Sternwarte, and the
    Radioastronomisches Institut der Universit\"at Bonn.}
  , Auf dem H\"ugel 71, D-53121 Bonn\\
  $^2$ Rhine Stellar Dynamics Network (RSDN)
}
\begin{document}
\maketitle
\begin{abstract}
  The ONC appears to be unusual on two grounds: 
  The observed constellation of the OB-stars of the entire Orion Nebula
  cluster and its Trapezium at its centre implies
  a time-scale problem given the
  age of the Trapezium, and an IMF problem for the whole
  OB-star population in the ONC. Given the estimated crossing
  time of the Trapezium, it ought to have totally dynamically decayed by now.
  Furthermore, by combining the lower limit of the ONC mass 
  with a standard IMF it emerges that the  ONC should have
  formed at least about 40 stars heavier than 5~M$_\odot$ while only ten
  are observed. Using $N$-body experiments we (i) confirm the
  expected instability of the trapezium and (ii) show
  that beginning with a compact OB-star configuration of
  about 40 stars the number of observed OB stars
  after 1 Myr within 1 pc radius and a compact trapezium
  configuration can both be reproduced. These two empirical constraints
  thus support our estimate of 40 initial OB stars in the cluster.
  Interestingly a more-evolved version of the ONC
  resembles the Upper Scorpius OB association.
  The $N$-body experiments
  are performed with the new C-code {\sc catena} by integrating the
  equations of motion using the chain-multiple-regularization
  method. In addition we present a new numerical formulation of the initial 
  mass function. 
  
\end{abstract}

\section{Introduction}
Of all O-stars 46 per cent and of 
all B-stars 4 per cent are runaways exceeding
30 km/s \citep{stone1991a}. Furthermore 
the binary fraction among runaway O-stars 
is around 10~\% \citep{gies1986a} while it is more than 50~\% in
young star clusters \citep{goodwin2006a}. This suggests that binaries
are involved in close dynamical encounters leading to stellar
ejections while the binary fraction among the ejected stars is
decreased. Indeed,  \citet{clarke1992a}
deduced using an analytical approach that massive stars must form 
in compact small-$N$ groups.
The decay of non-hierarchical 3,4,5-body systems with equal masses as well
as a mass spectrum has been investigated by \citet{sterzik1998a}.
They determined the spectrum of the remnant decay products but 
not the phase-space behaviour of compact few-body systems with time.
\citet*{hoogerwerf2000a} and \citet{hoogerwerf2001a}
were able to trace back the trajectories of some 
runaways to nearby associations. \citet*{ramspeck2001a}
determined the age and the calculated 
time-of-flight of early-type stars at high galactic
latitudes and concluded that they can have their
origin in the galactic disk. In the case of the 
runaways AE Aurigae and $\mu$ Columbae, which have 
spatial velocities greater than 100~km/s, in
combination with the binary $\iota$ Orionis,
\citet*{gualandris2004a} have shown
that the encounter of two binaries with high
eccentricities 2.5~Myr ago and in the co-moving vicinity of the current 
Orion Nebula Cluster (ONC) can reproduce the spatial configuration
observed today. 
The spatial distribution of field OB-stars can thus be understood 
qualitatively
using theoretical stellar-dynamical methods.  
But to obtain a more complete picture we need to study the details
and frequency of occurrence of energetic ejections from the acceleration
centres, namely the inner regions of young star clusters. 
\section{Motivating problems}
In the case of the Orion Nebula cluster two main discrepancies
concerning the properties of its OB-stars are found: 
\subsection{Existence of the Trapezium system}
Given the total mass of all four Trapezium stars \citep{hillenbrand1997a}
of 88.4~M$_\odot$ and their occupying space of nearly 0.05~pc in diameter the
corresponding crossing time can be estimated, 
\begin{equation}
t_{\mathrm{cr}} = \sqrt{\frac{4\;R^3}{G\;M}}\;,
\end{equation}
to be about 13 Kyr (Tab. \ref{setup}), 
if the Trapezium is assumed to be a compact 
virialized subsystem of the ONC.
\citet{sterzik1998a} noted that most systems in their decay
analysis decay within dozens of crossing times.  
So the ONC-TS is expected to 
have totally decayed by now, if its age is about
1 Myr \citep{kroupa2004a}. 
\begin{table}
  \caption{Setup data for all three models}
  \label{setup}
  \begin{tabular}{lcccrr}
    \hline
    model & stars & $\frac{M_\mathrm{tot,OB}}{M_\odot}$ &
    $\frac{\sigma_{\mathrm{3D}}}{\mathrm{km/s}}$ & 
    $\frac{t_\mathrm{cr}}{\mathrm{Kyr}}$
    &runs\\ 
    \hline
    4-body & $\Theta^1$ in Tab. \ref{stars}& 88.4 & 3.9 & 12.6 & 1000\\
    10-body & all in Tab. \ref{stars}& 167.2 & 5.4 & 9.2 & 1000\\
    40-body & 4 $\times$ Tab. \ref{stars}& 668.8 & 10.7 & 4.7 & 1000\\
    \hline
  \end{tabular}
  
  \medskip
  Specification of the N-body systems. $M_\mathrm{tot}$ is the total mass in 
  $\mathrm{M}_\odot$.
  The velocity dispersion $\sigma$
  is calculated from the total mass placed within the initial radius 
  of 0.025~pc corresponding to the
  ONC-TS size if virial equilibrium is assumed.
  $t_{cr}$ is the crossing time, and runs are
  the total number of integrated configurations.
\end{table}
\subsection{The number of OB stars}
\label{sec_ob_number}
The virial mass of the ONC is measured to be nearly 4500~M$_\odot$
but only about 1800~M$_\odot$ is visible in stellar mass
\citep{hillenbrand1998a} while cluster-formation models that match
the ONC suggest that it may have formed with 10$^4$~stars plus brown
dwarfs and that it is expanding now resembling a Pleiades type cluster
embedded in an expanding association
to a remarkable degree after 100~Myr \citep{kroupa2001b}. 
The observed mass of all stars heavier than 5~M$_\odot$
is 167~M$_\odot$. If the canonical IMF (App. \ref{finding-N-OB}) 
is normalized such that 1633~M$_\odot$ are
contained in the mass interval ranging from 0.01 up to 5~M$_\odot$ 
and for three different physically
possible upper stellar mass limits, $m_{\mathrm{max}*}$, 
of 80, 150~M$_\odot$ or $+\infty$ 
\citep{weidner2004a,oey2005a,figer2005a,koen2006a}, and different IMF-slopes 
above 1~M$_\odot$
the corresponding maximum stellar mass $m_{\mathrm{max}}$
and the expected number of OB-stars formed in the ONC 
can be calculated (Tab. \ref{OB-star-number}). 
The calculations are based on the
IMF by \citet{kroupa2001a} 
but with a numerically more convenient description 
\footnote[1]{A utility-IMF package and {\sc catena} including a full
documentation can be downloaded 
from the AIfA-webpage: http://www.astro.uni-bonn.de/}
(App. \ref{imf}).
 
Given the values in Tab. \ref{OB-star-number}  
the ONC should have formed about 38 OB-stars assuming the IMF to be
canonical ($\alpha_3 = 2.35$). But in a thorough survey of the ONC 
\citet{hillenbrand1997a} lists only 10 stars weighting more than 5~M$_\odot$
(Tab. \ref{stars}). 
9 of these 10 OB stars are located within a projected sphere 
of 1 pc around the
Trapezium system. The remaining B star (5.7~M$_\odot$) is placed
approximately 2.3~pc away from the Trapezium. 7 OB stars, including the three
most massive stars, are located within 0.5~pc around the Trapezium 
in projection.
If the IMF is steepened above 1~M$_\odot$ 
to $\alpha_3 = 2.7$
the number of expected OB-stars decreases down to 18. But the 
expected maximum stellar mass also decreases down to 
$m_\mathrm{max} = 28$ M$_\odot$,
whereas two observed stars are heavier. The existence of these stars
suggests that the IMF was
indeed normal. Note that the time-scale problem would persist even
if we allow $\alpha_3 = 2.7$. Because the IMF seems to be universal
\citep{kroupa2002a} a significant deviation from the calculated number
of 38 stars using the canonical IMF should not be expected.

As a check the number of stars heavier than
5~M$_\odot$ can also be estimated by normalising the canonical IMF
to the number of stars in the mass range 1--2~M$_\odot$ in the cluster.
Using  the stellar sample of \citet{hillenbrand1997a}, 
the number of stars heavier than 5~M$_\odot$ can be derived from 
the number of stars between 1 and 2~M$_\odot$ (70) and noting 
that in this mass regime 
only the non-embedded sources are listed. These amount to approximately half
of all stars \citep{hillenbrand1997a}. 
Thus, 26 OB-stars are expected to have formed in the ONC. 
The total
mass derived from this  mass regime is 1404~M$_\odot$, 22~\% less 
than the total
estimated mass used above. Given this uncertainty (13--38 stars heavier than 
5~M$_\odot$), we perform computations with 10 and 40 stars. As will
become apparent below, 40 OB stars are our preferred value.

Furthermore, if stars are drawn randomly 
from a universal IMF, the number of stars
heavier than 5~M$_\odot$ may not be the expectation value of 38.
The number can be smaller.  To estimate the probability that less than 
$k$ of $n$ stars have masses less than 5~M$_\odot$, drawing stars from 
an IMF has to be 
interpreted as a Bernoulli experiment: For the mass of the ONC,  
$M_\mathrm{ONC}$, the total number of stars, $n_\mathrm{tot}$, and the number
of stars, $n_\mathrm{>5}$, heavier than 5~M$_\odot$ can be calculated. 
If one star is drawn
from the IMF, the probability to get a star heavier than 5~M$_\odot$ is
\begin{equation}
p = n_\mathrm{> 5}/n_\mathrm{tot}\;\;\;.
\end{equation}
This experiment is repeated $n_\mathrm{tot}$ times. So the probability
to have a star heavier than 5~M$_\odot$ $k$ times  is
given by the Bernoulli-distribution,
\begin{equation}
p(k) = \left(n_\mathrm{tot}\atop k\right)p^{k}(1-p)^{n_\mathrm{tot}-k}\;\;\;.
\end{equation}
Because the event probability is small and the number of experiments
large the Poissonian limit can be applied. The probability is approximately
\begin{equation}
p(k) = \frac{\mu^k}{\mu!}e^{-\mu}\;\;\;,
\end{equation}
where $\mu = p \;n_\mathrm{tot} = n_\mathrm{>5}$.
So the total probability to get $k$ or fewer OB-stars is
\begin{equation}
P(\le k) = \sum_{i=0}^{i=k} p(i)\;\;\;.
\end{equation}
The probability to get 20, 10 or fewer OB-stars for two different
ONC-masses and two different physically possible upper stellar mass
limits is given  in Tab. \ref{IMF_deviation}. It is extremely
unlikely that only ten stars have formed in the ONC if the IMF is
universal.

We note that the same argument can be applied to a more-evolved 
population:
In an exploration of the full stellar population of the 
Upper Scorpius OB association, \citet{preibisch2002a} determined
a total stellar mass of $2060\;\mathrm{M}_\odot$ covering a volume
of 35~pc in diameter. For the supernova progenitor they deduced
a mass of $\approx 40-60\;\mathrm{M}_\odot$. An IMF steeper than 2.3
in the regime of massive stars would not have lead to the formation 
of such a massive star in the young star-cluster-stage of the 
Upper Scorpius OB association 5~Myr ago for this mass of 
$2060\;\mathrm{M}_\odot$ \citep{weidner2006a}. 
This further supports that the IMF may not be
steeper than 2.3 for massive stars.
\citet{preibisch2002a} listed 19 stars heavier than 5~M$_\odot$.
This is approximately half of the expected number of formed stars
more massive than 5~M$_\odot$ and constitutes the same problem as for the
ONC due to similar initial cluster masses. Therefore, it can be argued
that O and B stars may have been ejected from their star forming region very 
early after their formation.

\begin{table}
\caption{The number of expected OB stars and maximum stellar mass
  in the Orion Nebula cluster.}
\label{OB-star-number}
\begin{tabular}{cccccc}
&&\multicolumn{3}{c}{}&\\
\hline
$\alpha_3$&$m_{\mathrm{max}*}/\mathrm{M}_\odot:$&$+\infty$&150&80&obs.\\
\hline
2.35&$m_\mathrm{max}/\mathrm{M}_\odot$& 76.4& 59.5& 46.8&45.7\\
2.35&$N_{>5}$& 38.7&  38.3&  37.8&10\\
2.35&$M_\mathrm{tot}/\mathrm{M}_\odot$& 2103 & 2076& 2048& 1800\\
2.7&$m_\mathrm{max}/\mathrm{M}_\odot$& 28.6& 27.7& 26.1&45.7\\
2.7&$N_{>5}$& 18.4& 18.4&  18.3&10\\
2.7&$M_\mathrm{tot}/\mathrm{M}_\odot$& 1799& 1797& 1794& 1800\\
\hline
\end{tabular}

\medskip
Observed \citep{hillenbrand1997a} and expected 
maximum stellar mass ($m_\mathrm{max}$), 
number of stars more massive than 5 M$_\odot$ ($N_{>5}$),
and  total initial mass ($M_\mathrm{tot}$)
for the Orion Nebula Cluster
in dependence of three different physically possible 
upper stellar mass limits, $m_{\mathrm{max}*}$, and two different IMF-slopes, 
$\alpha_3$, for the
mass range from 1 $M_\odot$ up to $m_\mathrm{max}$. The mass range less 
than 1 M$_\odot$ is described by a Kroupa-IMF \citep{kroupa2001a}.
\end{table}

So two questions arise assuming the IMF is invariant: 
Why does the Trapezium still exist and where
are the missing OB-stars? 

\begin{table}
\caption{Probability of a deviation from a canonical IMF}
\label{IMF_deviation}
\begin{tabular}{ccccc}\hline
$M_\mathrm{ONC}/\mathrm{M}_\odot$& 1800 & 1800 &2200 &2200\\
$m_\mathrm{max*}/\mathrm{M}_\odot$& 80&150&80&150\\\hline
$n_\mathrm{tot}$ & 5209&5144 & 6337&6251\\
$\mu$& 33&33&41&41\\
$P(k\le10)$&$2.8\cdot 10^{-6}$&$2.8\cdot 10^{-6}$
&$7.6\cdot 10^{-9}$&$7.6\cdot 10^{-9}$\\
$P(k\le20)$&$1.0\cdot 10^{-2}$&$1.0\cdot 10^{-2}$
&$2.2\cdot 10^{-4}$&$2.2\cdot 10^{-4}$\\\hline
\end{tabular}

\medskip

The probability to draw 20 ($P(k\le 20$)), 10 ($P(k\le 10$))  
or fewer stars heavier than 5~M$_\odot$
from a Kroupa-IMF, the expectation value $\mu$ of the number of stars  
heavier than 5~M$_\odot$ and the total number of stars $n_\mathrm{tot}$
(equivalent to the number of repeated experiments) are calculated
for two different total cluster masses and two different physically 
possible upper stellar mass limits.
\end{table}

\section{Integrator}
To investigate the dynamics of the OB-stars
in the ONC we perform direct N-body integrations.
Because close encounters with high eccentricity
are very frequent in compact few-body systems
due to the grainy potential,
a multiple regularization technique is required
to reduce  energy errors and speed-up the calculations.
We combined in our own code ({\sc catena}\footnotemark[1]) the
very efficient {\sc chain}-regularization formalism 
developed by \citet{mikkola1990a,mikkola1993a} with 
an embedded Runge-Kutta
method of 8(9)-th order using a coefficient-set 
published by \citet{prince1981a}, instead of 
the 
Aarseth-{\sc chain}-Burlisch-Stoer integrator, 
to integrate
the regularized equations of motions.

Computer codes for studying
the dynamics of few body systems and star clusters or planetary systems
are available. A very valuable review of this kind of software industry
is given in \citet{aarseth1999a,aarseth2003a}. 
But, interestingly, there is a lack of software for calculating the dynamical
decay of systems with a few $\le$ $N$ $\le$ four dozen stars.
Our endeavour is to fill this gap by a sophisticated software tool 
allowing us to efficiently study the decay of hierarchical and non-hierarchical
configurations of some tens or hundreds stars down to the last 
remaining hard binaries or hierarchical higher-order multiple-stars,
with the long-term-aim of embedding {\sc catena} 
in a general-purpose $N$-body code.

An error analysis for the present application is
provided in Sec. \ref{section_error}.
\section{Initial conditions}
To address the questions mentioned above we investigate
three models which consist of the stars listed in Tab. 
\ref{setup}.

In the first model we study the stability of the 
actually observed Trapezium
system consisting of $\Theta^1$ A, $\Theta^1$ B, 
$\Theta^1$ C and $\Theta^1$ D
precisely.
In the second model, it is assumed that all currently observed
OB stars in the ONC (Tab. \ref{stars})
were initially in a compact configuration as a core
at the centre of the ONC, due either to mass segregation or
ab-initio.
In the third model we start with an OB core 
coming close to the expected number of 38. To find a
suitable set of stars, all presently observed OB stars are used four times
giving 40 stars (4 times Tab. \ref{stars}).

The compact settings of OB-stars are motivated by the outcome
of the analytical investigation by \citet{clarke1992a} that 
massive stars form in compact groups.
\citet{bonnell1998a} concluded that
the positions of massive stars in the Trapeziums cluster in Orion 
cannot be due to dynamical mass segregation, but must have formed 
in, or near, the centre of the cluster. 

For each of these three models 1000 configurations 
are created where the stars from Tab. \ref{setup} are uniformly 
distributed over a sphere with the compact Trapezium radius of 0.025 pc 
\citep{hillenbrand1997a}. The velocities
are drawn from a Gaussian distribution with a velocity dispersion resulting
from the virial theorem, 
\begin{equation}
\sigma = \frac{G\;M_{\mathrm{tot, OB}}}{R}
\end{equation}(Tab. \ref{setup}).
After this the velocities are  re-scaled slightly to ensure
initial virialisation.

This simple model does not include the rest of the ONC. To 
estimate its effect on the core decay the ratio of the 
internal and external forces can be calculated. The OB-star core
of radius $r$ consists of $n$ stars having the mean mass $m$. 
The gravitational force on one star is then
\begin{equation}
F_\mathrm{n} = G \frac{n m^2}{r^2}\;\;\;.
\end{equation}
The force exerted by the rest of the ONC on one star in the 
core can be estimated by the Plummer force 
\begin{equation}
F_\mathrm{pl} = G m M_\mathrm{pl} (r^2+b^2)^{-\frac{3}{2}} r\;\;\;,
\end{equation}
assuming the cluster can be represented reasonably well by a Plummer model,
which has been shown to be the case \citep{kroupa2001b}.
The resulting force ratio is
\begin{equation}
\Phi = 
\frac{F_\mathrm{n}}{F_\mathrm{pl}}=
\frac{n m}{M_\mathrm{pl}}
\left(1+\frac{b^2}{r^2}\right)^{\frac{3}{2}}\;\;\;.
\end{equation}
The mass $M_\mathrm{pl}$ is the cluster mass minus the
mass of the OB-stars.
The  resulting force ratios can be seen in Tab. \ref{force_ratios}.
The core dynamics is dominated by its self-gravitation.

The escape
velocities for the isolated core and the total Plummer sphere can also
be compared. Both are obtained from the conservation-of-energy-theorem.
The escape speed from the centre of the Plummer sphere, $v_\mathrm{e,pl}$, 
and the  escape speed from the surface of an isolated OB-core are given
by
\begin{equation}
v_\mathrm{e,pl} = \sqrt{
\frac{2 G m_\mathrm{cl}}{b}}
\;\;\;,\;\;\;
v_\mathrm{e,OB} = \sqrt{
\frac{2 G m_\mathrm{OB}}{r_0}}
\;\;\;,
\end{equation}
respectively, where
$r_0 = 0.025\;\mathrm{pc}$ is the initial radius of the OB-core.
For the 4- and 10-body model the escape speeds for the isolated model
and the true embedded situation are comparable. In the 40-body model
the escape speed from the core is dominated by the core itself.


\revised{A second issue associated with the cluster shell of
  low-mass stars is two-body relaxation between an OB-star and
  the low mass stars of the cluster. Energy may be transfered from the 
  OB-star core and ejected or evaporated OB-stars to the rest of the cluster.
  The relaxation time of the ONC is about 18~Myr \citep{kroupa2005a}.  The
  relaxation time for a heavy star is given by multiplying the
  relaxation time with the ratio of the mass of the most massive star
  and the mean stellar mass \citep{spitzer1987a}
  and describes the time-scale of a massive star 
  to sink towards the cluster centre,
  \begin{equation}
    t_\mathrm{relax,OB} \approx \frac{\bar m}{m_\mathrm{OB}}
    t_\mathrm{relax}\;\;\;, 
  \end{equation}
  where the average mass $\bar m$ of a star is 0.35 M$_\odot$ using a
  Kroupa-IMF. 
  The resulting energy transfer time-scale ranges from 0.14 Myr (45.7 M$_\odot$)
  up to 1.26 Myr (5 M$_\odot$), 
  thus being shorter or comparable to the time spanned by the simulations
  and therewith probably an important issue in our context,
  given the age of the ONC  $\approx$ 1~Myr.
  In the case of no equipartition instability,
  energy transfer stops after reaching energy equipartition,
  \begin{equation}
    \bar m <v^2> = m_\mathrm{OB} <v_\mathrm{OB}^2> \;\;\;,
  \end{equation}
  where $<v^2>$ ($<v_\mathrm{OB}^2>$) 
  is the mean square velocity of the mean-mass stars (OB-stars, respectively).
  Using a velocity dispersion of 2~km~s$^{-1}$ \citep{hillenbrand1997a}
  for the mean-mass stars,
  the relation above and the energy theorem it can be calculated
  that the velocity of a 5~M$_\odot$ (45~M$_\odot$) is low enough such that 
  the movement of the OB-stars is constrained to be within a radius of 
  0.026~pc (0.0084~pc). So the current observed 
  OB~core has an extension consistent
  with energy equipartition. Following \citet{heggie2003a} 
  the heavy stars are so concentrated that the lighter stars have been
  expelled from the core and they no longer have a significant role.
  This is also suggested by the observed deficit of low-mass stars
  in the core of the ONC \citep{hillenbrand1998a}.

  We conclude that the effect of two-body
  relaxation between low-mass stars and the OB-stars may be of 
  minor importance and that these simulations 
  suffice to demonstrate the time-scale
  problem of the ONC, and that the OB-star core-decay-model may explain
  the OB-star number problem of the ONC.
  While full-scale $N$-body calculations
  capture the entire relevant physics, our approximations allow us to
  compute a very large number of renditions (here 5000 in total) 
  which is necessary given the
  low frequency of massive stars. Future $N$-body calculations of
  individual set-ups will be used to check our results.}
  
\begin{table}
  \caption{Force ratio for the 4-, 10- and 40-body model}
  \label{force_ratios}
  \begin{tabular}{ccccccc}
    \hline
    $n$ & $M_\mathrm{OB}/\mathrm{M}_\odot$ &
    $m/\mathrm{M}_\odot$ & $\Phi$&$v_\mathrm{e,OB}$&
    $M_\mathrm{cl}/\mathrm{M}_\odot$&
    $v_\mathrm{e,pl}$\\
    \hline
    4 & 88.4 & 22.1  & 94.5 &5.6&1721.4&7.1\\
    10 & 167.2 & 16.7 & 178.6&7.7&1800.2&7.3\\
    40 & 668.8 & 16.7 & 714.2&15.4&2301.8&8.2\\
    \hline
  \end{tabular}

\medskip
$n$ is the number of stars the model consists of, $M_\mathrm{OB}$
(cf. Tab. \ref{setup})
is the total mass contained in the OB-stars, $m$ is the mean mass
of an OB-star, $\Phi$ is the resulting force ratio using a Plummer
mass of 1633~M$_\odot$. Given the observed core radius of the ONC
of about 0.19 pc \citep{hillenbrand1998a}
the related Plummer parameter of the ONC is about 0.3 pc.
$v_\mathrm{e,OB}$ is the escape speed in km/s 
from the 
surface of an OB-core with radius of 0.025 pc, $v_\mathrm{e,pl}$
is the escape speed in km/s from the centre of a Plummer sphere with mass
$M_\mathrm{cl} = 1633\;\mathrm{M}_\odot + M_\mathrm{OB}$.
\end{table}

\begin{table}
  \caption{Identity of the stars used
    in the three models (Tab. \ref{setup})}
  \label{stars}
  \begin{tabular}{lccr}
    \hline
    Name  & Parenago & SpT & $m/\mathrm{M}_\odot$\\
    \hline
    $\Theta^1$A & 1865 & O9V & 18.9 \\
    $\Theta^1$B & 1863 & B0V & 7.2 \\
    $\Theta^1$C & 1891 & O7V & 45.7\\
    $\Theta^1$D & 1889 & B0Vp & 16.6\\
    $\Theta^2$A & 1993 & O9V & 31.2\\
    $\Theta^2$B & 2031 & B1V & 12.0\\
    LP Ori & 1772 &      B2V & 7.2\\
    ---    & 1956 &      B3 & 6.4\\
    NU Ori & 2074 &      B1V & 16.3\\
    HD37115 & 2271&     B5V & 5.7\\
  \hline
\end{tabular}

\medskip
Stellar data for all OB-stars over 5 M$_\odot$
given by \citet{hillenbrand1997a}.
Spectral type after \citet{vanaltena1988a}.
\end{table}

\section{Finding Trapezium systems}
\label{sec_finding_TS}
We define a trapezium system to
consist of a few stars having pairwise distances of the same
order. Here the whole system is scanned to determine the
maximum number of stars in a configuration in which the pairwise
distances lie between two boundaries:
When studying the stability of the ONC Trapezium these
boundaries are 0.01 and 0.05 pc.
When studying the total OB-star distribution these
boundaries are 0 and 0.05 pc. 

\begin{figure}
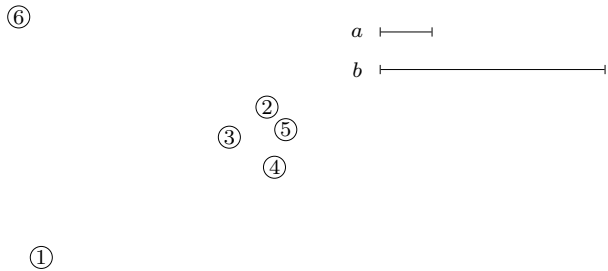

\pspicture*(-1,-1)(8.5,3)
\rput(-0.5,-0.5){1}
\rput(-0.8,2.7){6}
\rput(2.5,1.5){2}
\rput(2.0,1.1){3}
\rput(2.6,0.7){4}
\rput(2.75,1.2){5}
\psellipse[linewidth=0.2pt](-0.5,-0.5)(0.15,0.15)
\psellipse[linewidth=0.2pt](-0.8,2.7)(0.15,0.15)
\psellipse[linewidth=0.2pt](2.5,1.5)(0.15,0.15)
\psellipse[linewidth=0.2pt](2.0,1.1)(0.15,0.15)
\psellipse[linewidth=0.2pt](2.6,0.7)(0.15,0.15)
\psellipse[linewidth=0.2pt](2.75,1.2)(0.15,0.15)
\rput(3.7,2.5){$a$}
\psline[linewidth=0.3pt]{|-|}(4.0,2.5)(4.7,2.5)
\rput(3.7,2.0){$b$}
\psline[linewidth=0.3pt]{|-|}(4.0,2.0)(7.0,2.0)
\endpspicture
\caption{Illustration of the trapezium-finding routine
(see Sec.~{\ref{sec_finding_TS}} for explanation).}
\label{finding_TS}
\end{figure}

This procedure is illustrated in Fig. \ref{finding_TS}:
Consider a configuration consisting of six bodies. A table 
containing the pairwise distances is created. All subsets of particles
having a pairwise distance between two boundaries are determined.
Of all these subsets the one having the most members is the extracted
trapezium system. For the explicit example above, assume 
that a trapezium system of dimension $a$
is searched. The pairwise distances may be allowed to deviate by 20 per cent
from this dimension. Then the subsets found are: 
$\{2, 3\}$, $\{2, 4\}$, $\{2, 5\}$, $\{3, 4\}$, $\{3, 5\}$, $\{4, 5\}$,
$\{2, 3, 4\}$, $\{2, 3, 5\}$, $\{2, 4, 5\}$, $\{3, 4, 5\}$
, and $\{2, 3, 4, 5\}$. So the trapezium system consists of four
bodies. 

If a trapezium system of the dimension $b$ is searched all subsets
extracted from the distance table are: 
$\{1, 2, 6\}$, $\{1, 3, 6\}$, $\{1, 4, 6\}$, and $\{1, 5, 6\}$.
Four candidates for a trapezium system of dimension $b$ are found.
But they all have the same number of members. 

Based on this algorithm a set of bodies can contain 
no trapezium system of a certain 
dimension, or a trapezium system can have two or more members. But
it is not possible to find a trapezium system consisting only
of one body.

If the number of stars within a certain sphere is of interest 
the lower boundary must only be set to zero.  

The pairwise distances of the Trapezium stars in the ONC $\Theta 1$
lie approximately between 0.02 and 0.05 pc, whereby all of  the OB stars 
can be found
within nearly 1 pc radius around $\Theta 1$ \citep{hillenbrand1997a}. 
\section{Decay of OB-star cores}
All 3000 configurations are integrated over
2~Myr. 95\% of all runs have a relative energy error lower 
than $10^{-14}$ in the case  of the 4-body model, $10^{-12}$ 
for the 10-body model and $10^{-10}$ for the 40-body model. 
\subsection{Four-body model}
In the top-diagram of Fig. \ref{decay} the decay curve of a four-body 
trapezium consisting of the four stars of $\Theta^1$ Ori is
plotted. After 1~Myr
only 2.5 stars on average are found with a pairwise distance less than
0.05~pc, and 1 star on average is found with a 
pairwise distance between 0.05 and 0.01~pc, where in both cases 
the 3d curves lie slightly below the 2d-projection decay curve.
This demonstrates the time-scale problem pointed out above for the
observed Trapezium which is marked by the bold x. It can be argued
that these are only mean values which are gained from a number
distribution, and that a compact trapezium as is observed can survive for 1~Myr
with some probability. That this is not the case can be seen 
in Fig. \ref{histogram}, where the distribution of the member numbers
is plotted in a histogram. In 62.2~\% of all runs no stellar
configuration is found with a pairwise distance between 0.01 and 0.05~pc,
while in only 0.4~\% of all runs a trapezium of the observed size 
is found after 1~Myr.

On a first view this stands in contradiction to \citet{allen1974a},
which is the only known work about stability of trapezia systems.
Their result was that 63~\% of all trapezia  remain as a trapezia
after 30 crossing times ($\approx$~1~Myr). The reason is that
they used a slightly different definition of a trapezium system:
``let a multiple star system (of 3 or more stars)\ldots
if three or more such distances are of the same order of magnitude,
then the multiple system is of trapezium type.\ldots Two distances are of
the same order of magnitude, in this context, if their
ratio is greater than 1/3 but less than 3''. A four-body system satisfying
our definition of a trapezium system, i. e. all pairwise distances are of
the same order of magnitude, can evolve into a system of two singe stars and
one close binary. Then three distances are of the same order of magnitude,
i. e. the distance between the two single stars, and the distances between
each single star and one component of the binary. Their definition will
detect a trapezium system. Our algorithm also detects a trapezium system
but consisting only of three stars and having a different size than 
searched for. 
So the criterion by Allen and Poveda
does not take the number of members of the trapezium system into account nor
the size of the trapezium.
For a quantification of the stability of a four-body system, the crucial
point is
the number of stars the trapezium consists of.
The Allen-and-Poveda trapezia consist initially of six stars, and only one
of all 30 configurations (3.3~\%) retains  the initial size after 1~Myr, and 
consists finally only of four stars. Indeed in all their 30 runs binaries form
consisting preferentially of the two most massive stars, therewith
being quite consistent with our result.

\subsection{10-body model}
As in the four-body model the mean number of stars
within 0.05 and 1~pc is plotted in Fig. \ref{decay}.
The decay of an initial 
ten-body OB star core can neither reproduce a four-body trapezium
with a diameter of 0.05~pc nor the entire present-day ONC OB 
population (Fig. \ref{histogram}).  
\subsection{40-body model}
\begin{figure}
  \includegraphics[width=\figurewidth]{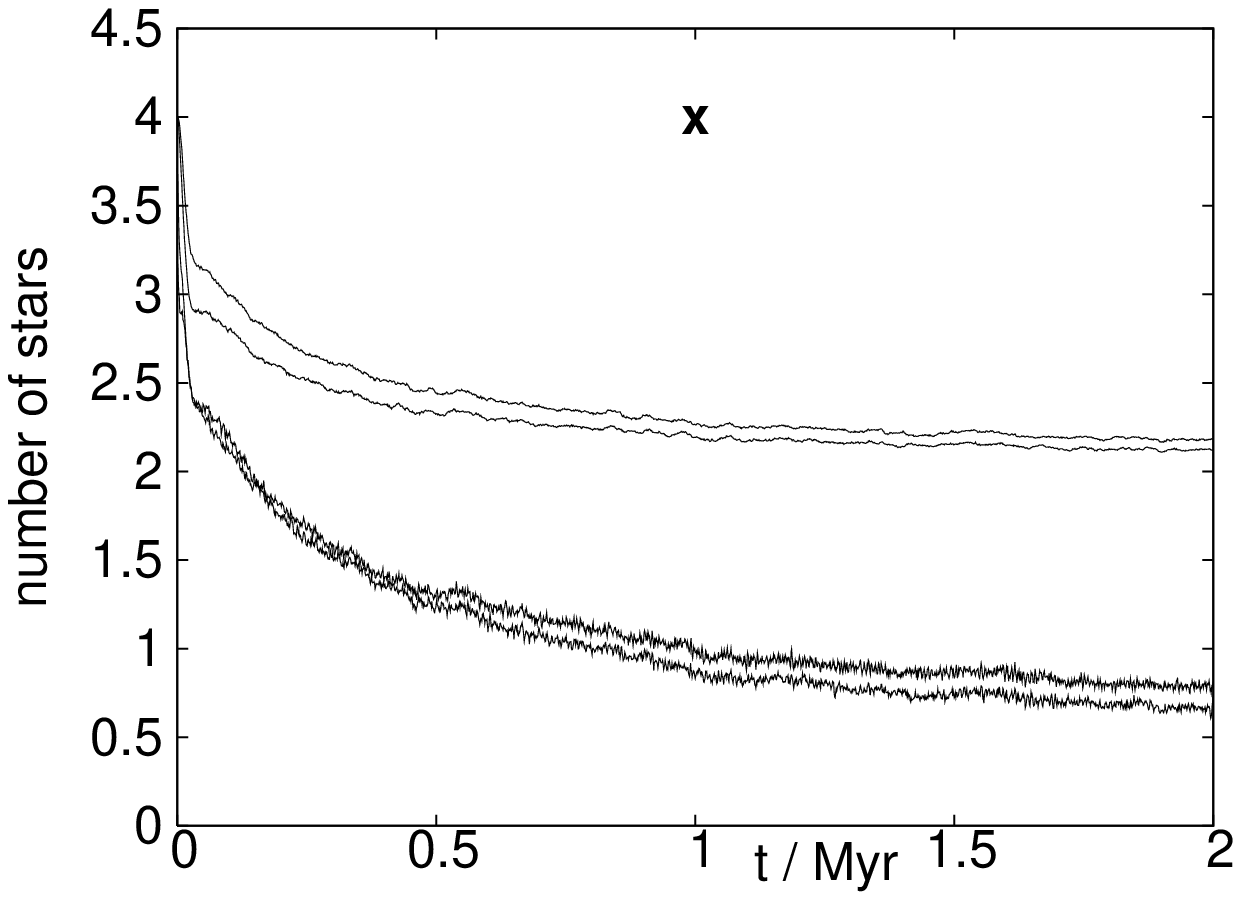}
  \includegraphics[width=\figurewidth]{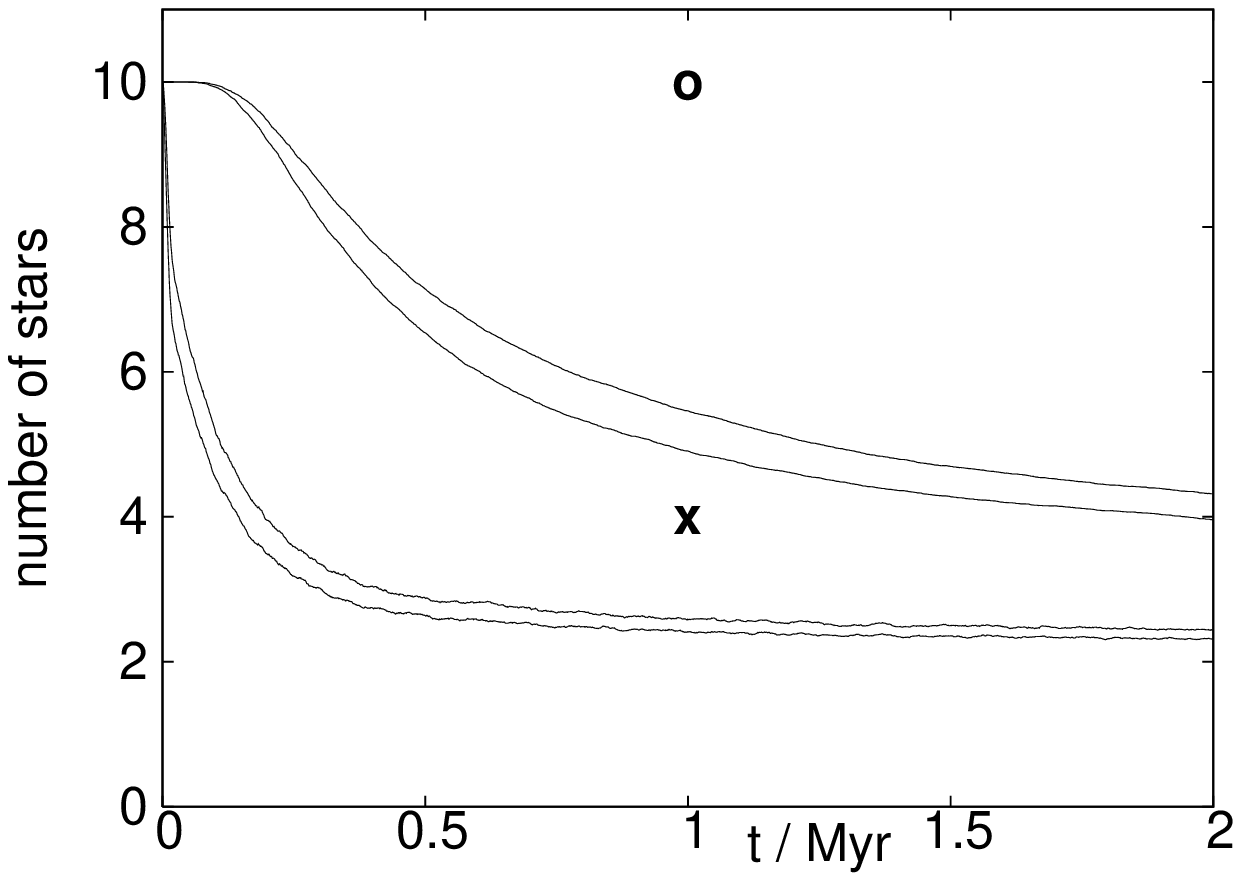}
  \includegraphics[width=\figurewidth]{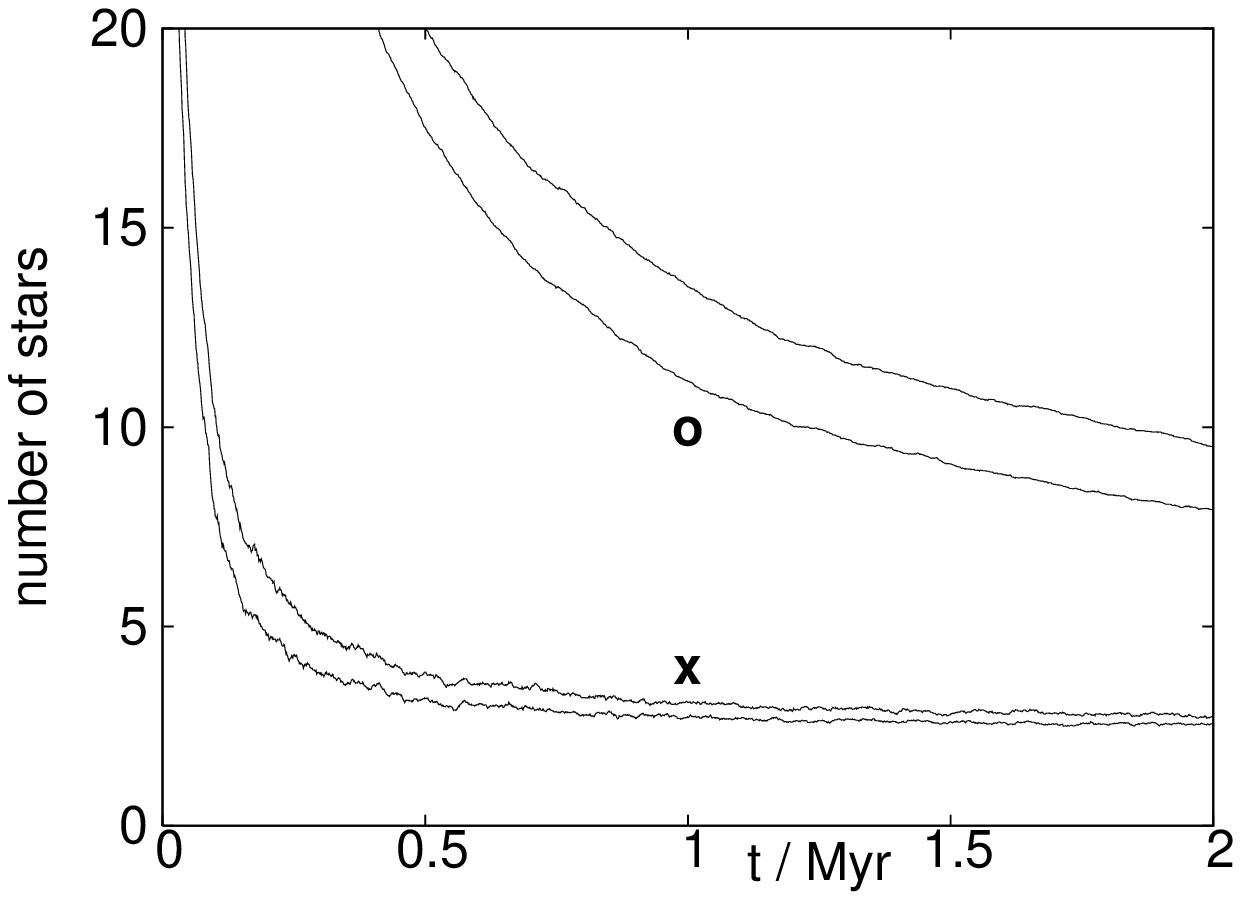}
  \caption{Decay curves for the $N$=4,10,40 models. 
    Plotted is the mean number of stars for  configurations
    with maximum member numbers having a certain pairwise 
    distance as a function of time. The bold x
    marks the position of the present ONC-TS with an assumed age of 1~Myr, 
    and the bold o marks the observed OB-stars in the ONC.
    {\it top}: Decay of the 4-body model where the pairwise
    distance lies (from bottom to top) between 0.01 and 0.05~pc (3d),
    between 0.01 and 0.05~pc (2d), below 0.05~pc (3d) and below 0.05~pc (2d).
    {\it middle}: Decay of the 10-body model where the pairwise
    distance lies (from bottom to top) below 0.05~pc (3d), below 0.05~pc (2d),
    below 1~pc (3d) and  below 1~pc (2d).
    {\it bottom}: Decay of the forty-body model where the pairwise
    distance lies (from bottom to top) below 0.05~pc (3d), below 0.05~pc (2d),
    below 1~pc (3d) and  below 1~pc (2d).
  }
\label{decay}
\end{figure}

If it is assumed that the ONC had an  initial OB-star content 
of nearly forty as expected
from the canonical IMF combined with the estimated 
stellar mass
of the ONC of about 2200~M$_{\odot}$ then the remaining number of OB stars
after 1~Myr within a sphere of 1~pc radius comes close to the observed value
of ten (Fig. \ref{decay}). 
The probability to observe a trapezium at an age of 0.5~Myr is 13.8~\%
but only 0.7~\% at an age of 1~Myr (Fig. \ref{histogram}). 
But counted together with the systems
containing more than four stars the probability to find
a compact trapezium increases to $0.7+2.9=3.6$\%. 
Note that the estimated ONC mass is comprised of the observed mass 
(1800~M$_\odot$) plus the estimated mass in missing 30 OB stars 
(400~M$_\odot$). The true initial ONC mass may have been about 4000~M$_\odot$
\citep{kroupa2001b}.

\begin{figure}
  \includegraphics[width=\columnwidth]{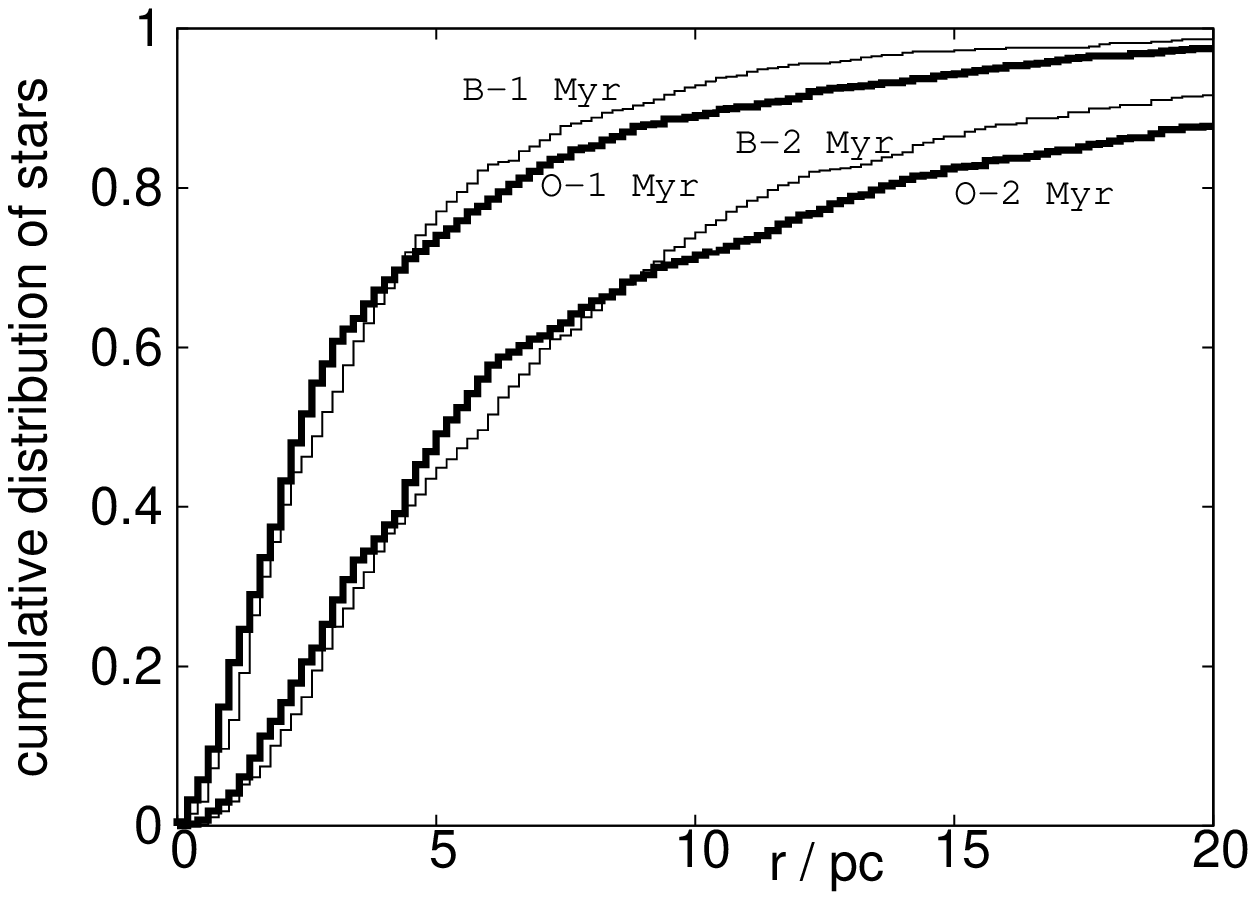}
  \includegraphics[width=\columnwidth]{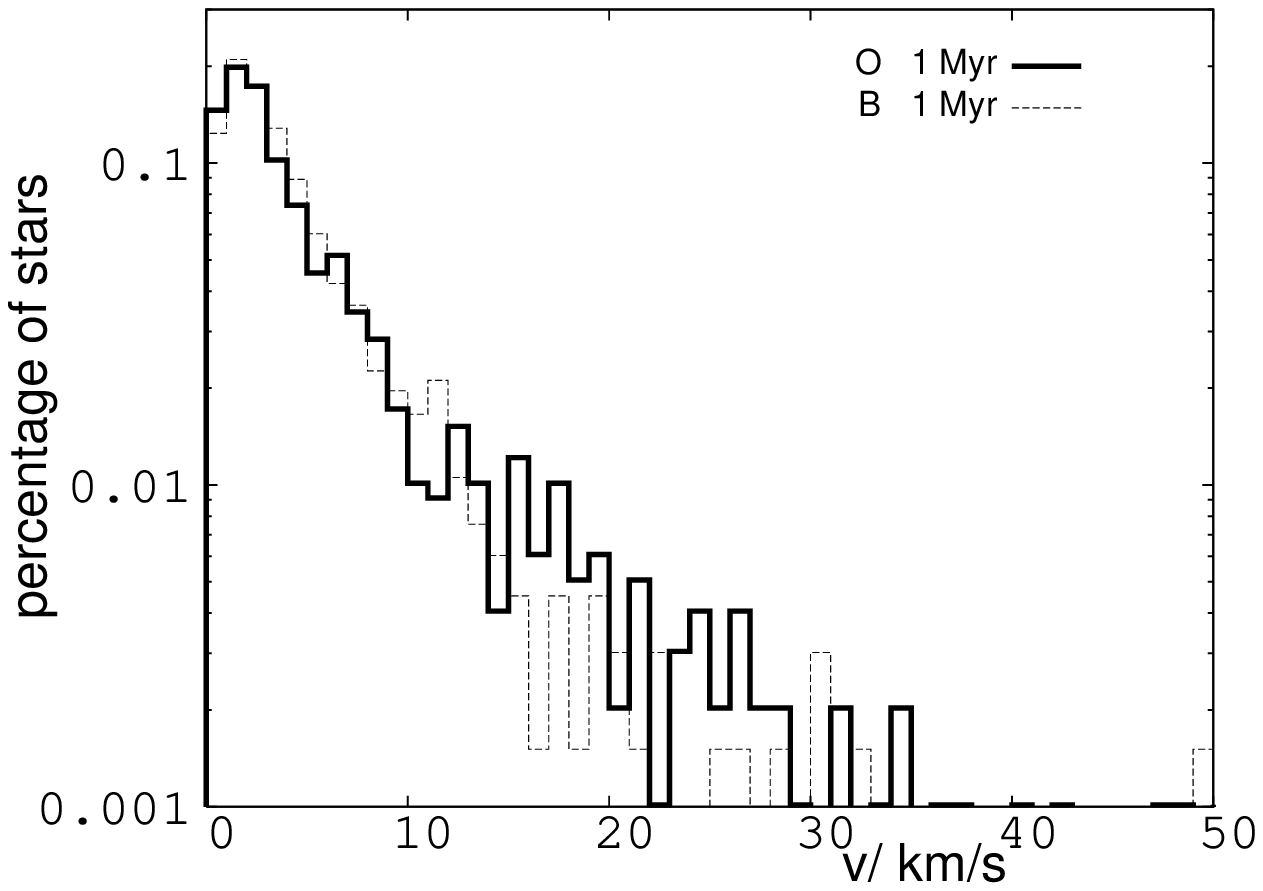}
\caption{
{\it top}: Cumulative spatial distribution of O and B stars
for the 40-body model 
after 1 and 2~Myr measured relative to the total centre of mass.
{\it bottom}: logarithmic Distribution of the velocities of O and B-stars for
the 40-body model after
1~Myr. Double stars are considered using their centre of mass velocity.}
\label{vr-distribution}
\end{figure}

In Fig. \ref{vr-distribution} the spatially cumulative 
and velocity distributions of the O and B
stars after 1 and 2~Myr are plotted. After 1~Myr more than 75~\% 
(30 of 40) of all
stars have larger distances to their common centre of mass than 
2~pc, and after 2~Myr 75~\% of all stars are more than 4~pc away 
from their centre of mass. Therefore only ten of forty stars heavier than
5~M$_\odot$ remain at the cluster centre as observed in the ONC.
It can be seen that more O stars than
B stars are at very large distances as well as more O stars than
B stars have very high velocities. This comes about because initially 
B stars tend to evaporate by energy redistribution rather than being ejected
by close encounters. Because O stars are heavier they form
tighter configurations than B stars and they are then involved in close 
binary interactions leading to high ejection velocities. This confirms the
result of \citet{clarke1992a} qualitatively. 

Given the spatial distribution of OB-stars after an evolution of 1 and
2~Myr, an extrapolation to an evolution age of 5 Myr predicts that 
nearly 50~\% of all OB-stars cover a volume of 25~pc 
in diameter which comes close to
the observed properties of the Upper Scorpius OB association
as pointed out in Sec. \ref{sec_ob_number}.

\begin{figure}

\vspace{0.2cm}
{\bf 4-body model}:

\psset{xunit=0.076923076\figurewidth}
\pspicture*(2.6,-0.4)(13.5,3)
\psset{yunit=0.001}
\psset{linewidth=0.01cm}
%
\rput(4.4,250){after 0.5~Myr:}
\psline{->}(6.8,-50)(13.2,-50)
\psline{->}(6.8,-50)(6.8,2500)
\rput(5.8,-250){$N_{\mathrm{stars}}$:}
\rput(5.8,2500){per-}
\rput(5.8,2300){centage}
\psline(7,0)(8,0)
\psline(7,452)(8,452)
\psline(7,0)(7,452)
\psline(8,0)(8,452)
\rput(7.5,652){45.2}
\rput(7.5,-250){0}
\psline(8.2,0)(9.2,0)
\psline(8.2,401)(9.2,401)
\psline(8.2,0)(8.2,401)
\psline(9.2,0)(9.2,401)
\rput(8.7,601){40.1}
\rput(8.7,-250){2}
\psline(9.4,0)(10.4,0)
\psline(9.4,137)(10.4,137)
\psline(9.4,0)(9.4,137)
\psline(10.4,0)(10.4,137)
\rput(9.9,337){13.7}
\rput(9.9,-250){3}
\psline(10.6,0)(11.6,0)
\psline(10.6,10)(11.6,10)
\psline(10.6,0)(10.6,10)
\psline(11.6,10)(11.6,10)
\rput(11.1,210){1.0}
\rput(11.1,-250){4}
\psline(11.8,0)(12.8,0)
\psline(11.8,0)(12.8,0)
\psline(11.8,0)(11.8,0)
\psline(12.8,0)(12.8,0)
\rput(12.3,200){0}
\rput(12.3,-260){$>4$}
%
%
%
\rput(4.4,1750){after 1.0~Myr:}
\psline(7,1500)(8,1500)
\psline(7,2122)(8,2122)
\psline(7,1500)(7,2122)
\psline(8,1500)(8,2122)
\rput(7.5,2322){62.2}
\psline(8.2,1500)(9.2,1500)
\psline(8.2,1807)(9.2,1807)
\psline(8.2,1500)(8.2,1807)
\psline(9.2,1500)(9.2,1807)
\rput(8.7,2007){30.7}
\psline(9.4,1500)(10.4,1500)
\psline(9.4,1567)(10.4,1567)
\psline(9.4,1500)(9.4,1567)
\psline(10.4,1500)(10.4,1567)
\rput(9.9,1767){6.7}
\psline(10.6,1500)(11.6,1500)
\psline(10.6,1500)(11.6,1504)
\psline(10.6,1500)(10.6,1504)
\psline(11.6,1504)(11.6,1504)
\rput(11.1,1704){0.4}
\psline(11.8,1500)(12.8,1500)
\psline(11.8,1500)(12.8,1500)
\psline(11.8,1500)(11.8,1500)
\psline(12.8,1500)(12.8,1500)
\rput(12.3,1700){0}
\endpspicture

\vspace{0.2cm}
{\bf 10-body model}:

\psset{xunit=0.076923076\figurewidth}
\pspicture*(2.6,-0.4)(13.5,3)
\psset{yunit=0.001}
\psset{linewidth=0.01cm}
%
\rput(4.4,250){after 0.5 Myr:}
\psline{->}(6.8,-50)(13.2,-50)
\psline{->}(6.8,-50)(6.8,2500)
\rput(5.8,-250){$N_{\mathrm{stars}}$:}
\rput(5.8,2500){per-}
\rput(5.8,2300){centage}
\psline(7,0)(8,0)
\psline(7,320)(8,320)
\psline(7,0)(7,320)
\psline(8,0)(8,320)
\rput(7.5,520){32.0}
\rput(7.5,-250){0}
\psline(8.2,0)(9.2,0)
\psline(8.2,525)(9.2,525)
\psline(8.2,0)(8.2,525)
\psline(9.2,0)(9.2,525)
\rput(8.7,725){52.5}
\rput(8.7,-250){2}
\psline(9.4,0)(10.4,0)
\psline(9.4,134)(10.4,134)
\psline(9.4,0)(9.4,134)
\psline(10.4,0)(10.4,134)
\rput(9.9,334){13.4}
\rput(9.9,-250){3}
\psline(10.6,0)(11.6,0)
\psline(10.6,21)(11.6,21)
\psline(10.6,0)(10.6,12)
\psline(11.6,21)(11.6,21)
\rput(11.1,221){2.1}
\rput(11.1,-250){4}
\psline(11.8,0)(12.8,0)
\psline(11.8,0)(12.8,0)
\psline(11.8,0)(11.8,0)
\psline(12.8,0)(12.8,0)
\rput(12.3,200){0}
\rput(12.3,-260){$>4$}
%
%
%
\rput(4.4,1750){after 1.0 Myr:}
\psline(7,1500)(8,1500)
\psline(7,1947)(8,1947)
\psline(7,1500)(7,1947)
\psline(8,1500)(8,1947)
\rput(7.5,2147){44.7}
\psline(8.2,1500)(9.2,1500)
\psline(8.2,1992)(9.2,1992)
\psline(8.2,1500)(8.2,1992)
\psline(9.2,1500)(9.2,1992)
\rput(8.7,2192){49.2}
\psline(9.4,1500)(10.4,1500)
\psline(9.4,1559)(10.4,1559)
\psline(9.4,1500)(9.4,1559)
\psline(10.4,1500)(10.4,1559)
\rput(9.9,1759){5.9}
\psline(10.6,1500)(11.6,1500)
\psline(10.6,1500)(11.6,1502)
\psline(10.6,1500)(10.6,1502)
\psline(11.6,1502)(11.6,1502)
\rput(11.1,1702){0.2}
\psline(11.8,1500)(12.8,1500)
\psline(11.8,1500)(12.8,1500)
\psline(11.8,1500)(11.8,1500)
\psline(12.8,1500)(12.8,1500)
\rput(12.3,1700){0}
\endpspicture
%
%

\vspace{0.2cm}
{\bf 40-body model}:

\psset{xunit=0.076923076\figurewidth}
\pspicture*(2.6,-0.4)(13.5,3)
\psset{yunit=0.001}
\psset{linewidth=0.01cm}
%
\rput(4.4,250){after 0.5 Myr:}
\psline{->}(6.8,-50)(13.2,-50)
\psline{->}(6.8,-50)(6.8,2500)
\rput(5.8,-250){$N_{\mathrm{stars}}$:}
\rput(5.8,2500){per-}
\rput(5.8,2300){centage}
\psline(7,0)(8,0)
\psline(7,14)(8,14)
\psline(7,0)(7,14)
\psline(8,0)(8,14)
\rput(7.5,214){1.4}
\rput(7.5,-250){0}
\psline(8.2,0)(9.2,0)
\psline(8.2,460)(9.2,460)
\psline(8.2,0)(8.2,460)
\psline(9.2,0)(9.2,460)
\rput(8.7,660){46.0}
\rput(8.7,-250){2}
\psline(9.4,0)(10.4,0)
\psline(9.4,352)(10.4,352)
\psline(9.4,0)(9.4,352)
\psline(10.4,0)(10.4,352)
\rput(9.9,552){35.2}
\rput(9.9,-250){3}
\psline(10.6,0)(11.6,0)
\psline(10.6,138)(11.6,138)
\psline(10.6,0)(10.6,138)
\psline(11.6,0)(11.6,138)
\rput(11.1,338){13.8}
\rput(11.1,-250){4}
\psline(11.8,0)(12.8,0)
\psline(11.8,36)(12.8,36)
\psline(11.8,0)(11.8,36)
\psline(12.8,0)(12.8,36)
\rput(12.3,236){3.6}
\rput(12.3,-260){$>4$}
%
%
%
\rput(4.4,1750){after 1.0 Myr:}
\psline(7,1500)(8,1500)
\psline(7,1572)(8,1572)
\psline(7,1500)(7,1572)
\psline(8,1500)(8,1572)
\rput(7.5,1772){7.2}
\psline(8.2,1500)(9.2,1500)
\psline(8.2,2119)(9.2,2119)
\psline(8.2,1500)(8.2,2119)
\psline(9.2,1500)(9.2,2119)
\rput(8.7,2319){61.9}
\psline(9.4,1500)(10.4,1500)
\psline(9.4,1773)(10.4,1773)
\psline(9.4,1500)(9.4,1773)
\psline(10.4,1500)(10.4,1773)
\rput(9.9,1974){27.3}
\psline(10.6,1500)(11.6,1500)
\psline(10.6,1507)(11.6,1507)
\psline(10.6,1500)(10.6,1507)
\psline(11.6,1500)(11.6,1507)
\rput(11.1,1707){0.7}
\psline(11.8,1500)(12.8,1500)
\psline(11.8,1529)(12.8,1529)
\psline(11.8,1500)(11.8,1529)
\psline(12.8,1500)(12.8,1529)
\rput(12.3,1729){2.9}
\endpspicture

\caption{Distribution of the remaining number of stars  
  in trapezium configurations having a pairwise stellar distance
  between 0.01 and 0.05~pc in 2-d
  projection after 0.5 and 1~Myr.
  {\it top}: 4-body model, {\it middle}: 
  10-body model, {\it bottom}: 40-body model.}
\label{histogram}
\end{figure}

\section{Error Analysis}
\label{section_error}
An important question concerning $N$-body
simulations is whether we can trust them. This 
question arises from the fact that 
the basic process for the decay of $N$-body
systems are close and highly eccentric encounters of stars
with a consequent redistribution of
energy. These kind of encounters are the most
important sources for orbit-errors. 
Due to the exponential instability the numerical and true solution
deviate increasingly with time.
\citep{goodman1993a,heggie2003a}

As there exists no
general analytic solution for systems with more
than 2 bodies the integrals of
motion have to be checked for conservation to control the integration.
The most sensitive quantity is the total energy.
Smaller step sizes reduce the amount of the error but
prolong the duration of the integration, while step-sizes
that are too small lead to the accumulation of roundoff errors. So a
balance between efficiency and accuracy needs to be found.

\citet{dejonghe1986a} determined the accuracy of the integration
of 14  small-$N$ configurations using the time-reversal-test compared 
with energy conservation, leading to the
conclusion \citep{aarseth2003a} that
using the energy error only is questionable
for establishing exact integrations.

When using the statistical approach, high accuracy is not essential
to obtain meaningful results, provided the sample is sufficiently large
\citep{aarseth2003a}. \citet{valtonen1974a} determined that the 
distributions of eccentricity, terminal escape velocity and life-time
in 200 3D experiments did not show any clear accuracy dependence for
a relative energy error range from $5\cdot 10^{-4}$ to $3\cdot 10^{-2}$. 

Therefore, despite the fact that the orbits are completely wrong
after many crossing times, the statistical outcome from many equivalent
$N$-body experiments is reliable.

\begin{figure}
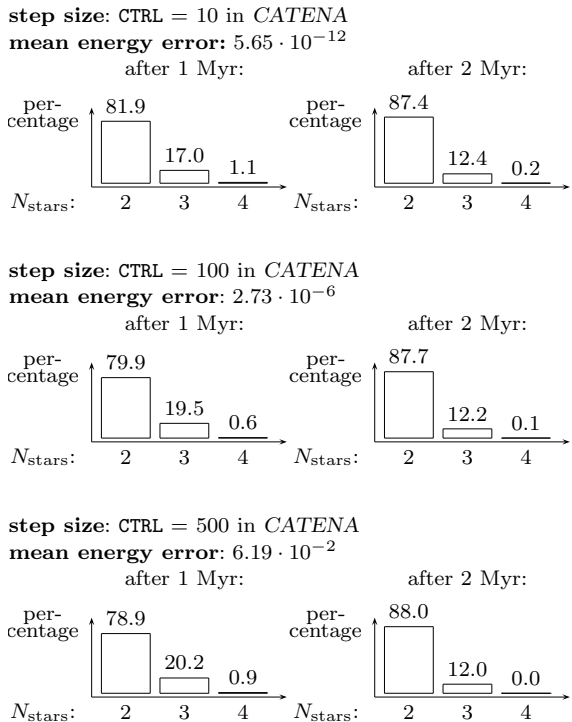

  {\bf step size}: {\tt CTRL} = 10 in {\sl CATENA}\newline
  {\bf mean energy error:} $5.65\cdot 10^{-12}$
  
  \psset{xunit=0.076923076\figurewidth}
  \pspicture*(-0.7,-0.4)(13,1.7)
  \psset{yunit=0.001}
  \psset{linewidth=0.01cm}
  \rput(2.9,1500){after 1~Myr:}
  \psline{->}(1,-50)(5,-50)
  \psline{->}(1,-50)(1,1000)
  \rput(0,-250){$N_{\mathrm{stars}}$:}
  \rput(0,1000){per-}
  \rput(0,800){centage}
  \psline(1.2,0)(2.2,0)
  \psline(1.2,819)(2.2,819)
  \psline(1.2,0)(1.2,819)
  \psline(2.2,0)(2.2,819)
  \rput(1.7,1019){81.9}
  \rput(1.7,-250){2}
  \psline(2.4,0)(3.4,0)
  \psline(2.4,170)(3.4,170)
  \psline(2.4,0)(2.4,170)
  \psline(3.4,0)(3.4,170)
  \rput(2.9,370){17.0}
  \rput(2.9,-250){3}
  \psline(3.6,0)(4.6,0)
  \psline(3.6,11)(4.6,11)
  \psline(3.6,0)(3.6,11)
  \psline(4.6,0)(4.6,11)
  \rput(4.1,211){1.1}
  \rput(4.1,-250){4}
  %
  \psline{->}(6.8,-50)(10.8,-50)
  \psline{->}(6.8,-50)(6.8,1000)
  \rput(5.8,-250){$N_{\mathrm{stars}}$:}
  \rput(5.8,1000){per-}
  \rput(5.8,800){centage}
  \rput(8.7,1500){after 2~Myr:}
  \psline(7,0)(8,0)
  \psline(7,874)(8,874)
  \psline(7,0)(7,874)
  \psline(8,0)(8,874)
  \rput(7.5,1074){87.4}
  \rput(7.5,-250){2}
  \psline(8.2,0)(9.2,0)
  \psline(8.2,124)(9.2,124)
  \psline(8.2,0)(8.2,124)
  \psline(9.2,0)(9.2,124)
  \rput(8.7,324){12.4}
  \rput(8.7,-250){3}
  \psline(9.4,0)(10.4,0)
  \psline(9.4,2)(10.4,2)
  \psline(9.4,0)(9.4,2)
  \psline(10.4,0)(10.4,2)
  \rput(9.9,202){0.2}
  \rput(9.9,-250){4}
  \endpspicture
  \vspace{0.5cm}
  {\bf step size}: {\tt CTRL} = 100 in {\sl CATENA}\newline
  {\bf mean energy error}: $2.73\cdot 10^{-6}$

  \psset{xunit=0.076923076\figurewidth}
  \pspicture*(-0.7,-0.4)(13,1.7)
  \psset{yunit=0.001}
  \psset{linewidth=0.01cm}
  \psline{->}(1,-50)(5,-50)
  \psline{->}(1,-50)(1,1000)
  \rput(0,-250){$N_{\mathrm{stars}}$:}
  \rput(0,1000){per-}
  \rput(0,800){centage}
  \rput(2.9,1500){after 1~Myr:}
  \psline(1.2,0)(2.2,0)
  \psline(1.2,799)(2.2,799)
  \psline(1.2,0)(1.2,799)
  \psline(2.2,0)(2.2,799)
  \rput(1.7,999){79.9}
  \rput(1.7,-250){2}
  \psline(2.4,0)(3.4,0)
  \psline(2.4,195)(3.4,195)
  \psline(2.4,0)(2.4,195)
  \psline(3.4,0)(3.4,195)
  \rput(2.9,395){19.5}
  \rput(2.9,-250){3}
  \psline(3.6,0)(4.6,0)
  \psline(3.6,6)(4.6,6)
  \psline(3.6,0)(3.6,6)
  \psline(4.6,0)(4.6,6)
  \rput(4.1,206){0.6}
  \rput(4.1,-250){4}
  %
  \psline{->}(6.8,-50)(10.8,-50)
  \psline{->}(6.8,-50)(6.8,1000)
  \rput(5.8,-250){$N_{\mathrm{stars}}$:}
  \rput(5.8,1000){per-}
  \rput(5.8,800){centage}
  \rput(8.7,1500){after 2~Myr:}
  \psline(7,0)(8,0)
  \psline(7,877)(8,877)
  \psline(7,0)(7,877)
  \psline(8,0)(8,877)
  \rput(7.5,1077){87.7}
  \rput(7.5,-250){2}
  \psline(8.2,0)(9.2,0)
  \psline(8.2,122)(9.2,122)
  \psline(8.2,0)(8.2,122)
  \psline(9.2,0)(9.2,122)
  \rput(8.7,322){12.2}
  \rput(8.7,-250){3}
  \psline(9.4,0)(10.4,0)
  \psline(9.4,1)(10.4,1)
  \psline(9.4,0)(9.4,1)
  \psline(10.4,0)(10.4,1)
  \rput(9.9,201){0.1}
  \rput(9.9,-250){4}
  \endpspicture
  \vspace{0.5cm}
  {\bf step size}: {\tt CTRL} = 500 in {\sl CATENA}\newline
  {\bf mean energy error}: $6.19\cdot 10^{-2}$

  \psset{xunit=0.076923076\figurewidth}
  \pspicture*(-0.7,-0.4)(13,1.7)
  \psset{yunit=0.001}
  \psset{linewidth=0.01cm}
  \psline{->}(1,-50)(5,-50)
  \psline{->}(1,-50)(1,1000)
  \rput(0,-250){$N_{\mathrm{stars}}$:}
  \rput(0,1000){per-}
  \rput(0,800){centage}
  \rput(2.9,1500){after 1~Myr:}
  \psline(1.2,0)(2.2,0)
  \psline(1.2,789)(2.2,789)
  \psline(1.2,0)(1.2,789)
  \psline(2.2,0)(2.2,789)
  \rput(1.7,989){78.9}
  \rput(1.7,-250){2}
  \psline(2.4,0)(3.4,0)
  \psline(2.4,202)(3.4,202)
  \psline(2.4,0)(2.4,202)
  \psline(3.4,0)(3.4,202)
  \rput(2.9,402){20.2}
  \rput(2.9,-250){3}
  \psline(3.6,0)(4.6,0)
  \psline(3.6,9)(4.6,9)
  \psline(3.6,0)(3.6,9)
  \psline(4.6,0)(4.6,9)
  \rput(4.1,209){0.9}
  \rput(4.1,-250){4}
  %
  \psline{->}(6.8,-50)(10.8,-50)
  \psline{->}(6.8,-50)(6.8,1000)
  \rput(5.8,-250){$N_{\mathrm{stars}}$:}
  \rput(5.8,1000){per-}
  \rput(5.8,800){centage}
  \rput(8.7,1500){after 2~Myr:}
  \psline(7,0)(8,0)
  \psline(7,880)(8,880)
  \psline(7,0)(7,880)
  \psline(8,0)(8,880)
  \rput(7.5,1080){88.0}
  \rput(7.5,-250){2}
  \psline(8.2,0)(9.2,0)
  \psline(8.2,120)(9.2,120)
  \psline(8.2,0)(8.2,120)
  \psline(9.2,0)(9.2,120)
  \rput(8.7,320){12.0}
  \rput(8.7,-250){3}
  \psline(9.4,0)(10.4,0)
  \psline(9.4,0)(10.4,0)
  \psline(9.4,0)(9.4,0)
  \psline(10.4,0)(10.4,0)
  \rput(9.9,200){0.0}
  \rput(9.9,-250){4}
  
  \endpspicture
  \caption{Error analysis using the example of the four-body decay
    determining the maximum number of members in configurations
    with a pairwise distance below 0.05~pc in 3d after 1~Myr
    ({\it left column}) and 2~Myr ({\it right column}) 
    for three different step
    size parameters resulting in mean energy errors of $5.65\cdot 10^{-12}$
    ({\it top}), $2.73\cdot 10^{-6}$ ({\it middle}) 
    and $6.19\cdot 10^{-2}$ 
    ({\it bottom diagram}).}
  \label{error}
\end{figure}

To test what value of energy error is acceptable
we run all 1000 four-body configurations three times
with different step-size parameters.
The resulting mean energy errors are 
$5.65\cdot 10^{-12}$, $2.73\cdot 10^{-6}$ 
and $6.19\cdot 10^{-2}$ with increasing step-size
parameter (Fig. \ref{error}). 

To compare the statistical error with the numerical error
we interpret this analysis as a series of Bernoulli-experiments.
One experiment is the determination of a trapezium consisting of
$N_\mathrm{stars}$ stars after a certain time $T$. The outcome can
be {\it yes} with probability $p$ and {\it no} with the probability $1-p$.
This experiment is repeated $n = 1000$ times. Therefore the probability
to get $k$ times the event {\it yes} is given by
\begin{equation}
P(k) = \left(n\atop k\right)\;p^k\;\left(1-p\right)^{n-k}.
\end{equation}
The event probability is approximated by the mean probability
\begin{equation}
p = \bar p = (k_1+k_2+k_3/3)/n = \bar k / n,
\end{equation}
where $k_\mathrm{i}$ is the number of events in each of the three
experiment series. The variance of the Bernoulli-distribution
is given by 
\begin{equation}
\Delta k^2 = n\;p-p^2.
\end{equation}
The corresponding probability $p$, number of experiments $n$, mean number
of events $\bar k$, variance $\sqrt{\Delta k^2}$ and one sigma errors
$\Delta \%$ for the histograms in Fig. \ref{error} are given 
in Tab. \ref{bernoulli-errors}. As an example we consider trapezia 
consisting of 2 bodies after 1~Myr. The mean number of runs having
a trapezium of 2 bodies after 1~Myr is 802 out of 1000 (80.2~\%).
The one sigma variance is 2.8~\%. So all three experiments (81.9~\%, 
79.9~\% and 78.9~\%) lie within 1~sigma around the mean value 
(80.2~\% $\pm$ 2.8~\% = 77.4~\% -- 80.2~\%). So if a numerical error
arises from the different choice of the step-size parameter it
is not larger than the statistical error.
We conclude that we can trust these $N$-body
simulations. 

From the virial theorem
\begin{equation}
E= V/2,
\end{equation}
where $E$ is the total energy and $V$ is the potential energy, 
the relative energy error is
\begin{equation}
\Delta E/E \approx \Delta V/V,
\end{equation}
and with
\begin{equation}
V = -G\;\frac{M^2}{R},
\end{equation}
\begin{equation}
\Delta E/E \approx \Delta R/R,
\end{equation}
follows. This means that uncertainty in the pairwise distances is
of the order of the mean energy error.
As the energy errors are smaller than $6\cdot 10^{-2}$
the distance errors have only a very slight effect
on the number statistics.
\begin{table}
  \caption{One sigma errors for the four-body decay (Fig. \ref{error}) 
    interpreted as Bernoulli-experiment.}
  \label{bernoulli-errors}
  \begin{tabular}{ccccccc}\hline
    $T$/Myr & 1 & 1 & 1 & 2 & 2 & 2\\
    $N_\mathrm{stars}$ & 2 & 3 & 4 & 2 & 3 & 4\\
    n& 1000 & 1000 & 1000 & 1000 & 1000 & 1000 \\
    $\bar{k}$ &802.3& 18.9& 8.7& 877.0& 122.0& 1.0\\
    $p$ & 0.802& 0.189& 0.087& 0.877& 0.122& 0.010\\
    $\sqrt{\Delta k^2}$ & 28.3 & 13.7& 9.3& 29.6& 11.0&3.2\\
    $\Delta \%$ & 2.8& 1.4& 0.9& 3.0& 1.1& 0.3\\
    \hline
  \end{tabular} 
  
  \medskip
  Resulting variances for the error histograms in Fig. \ref{error}
  if the search for a trapezium consisting of $N_\mathrm{stars}$
  is interpreted as a Bernoulli-experiment. 
\end{table}
\section{Conclusions}
We have pointed out that in the case of the Orion Nebula cluster
two main problems exist: Its short decay-time implying the question as
to why it exists, and the significant number of missing OB stars implying
either that they have been lost if the IMF is invariant, or that the IMF
had a highly non-standard $\alpha_3 > 2.7$ which is unlikely because no
other stellar population in a cluster
with such an IMF is known to exist and the observed
most-massive star in the ONC is significantly larger than that expected 
for this non-standard IMF. 
It is extremely unlikely (3 out of 1000 cases) that a compact Trapezium
system consisting of four stars can survive for more than 1~Myr.
The assumption that the initial number of OB stars was about
40 increases the probability to observe a Trapezium system 
after 1~Myr (36 out of 1000 cases). 
We infer that the ONC Trapezium system could
be an OB-star core in its final stage of decay.

This scenario is supported by the fact that the spatial distribution
of these forty OB-stars obtained from the numerical simulations 
after 5~Myr comes close to the 
observed spatial distribution of the OB-stars in the Upper
Scorpius OB association which is assumed, using the total mass, 
to have had the same young star-cluster progenitor as the ONC. 

Given its total mass the ONC indeed ought to have
four times as many OB stars than are observed, namely 40. 
This suggests that about 30 OB stars may have been expelled from
the ONC if the IMF is invariant. Starting
with an initial number of OB stars of about 40, stars are ejected
due to three-body encounters, so that this model matches
the observed number of OB stars in the ONC and the ONC Trapezium.

As it has been shown that the probability that only 10 OB-stars have
formed in the ONC is very small, the missing stars must be somewhere.
After 1~(2)~Myr 90 (70)~\% of all OB-stars should be found within 
a 10~pc radius
around the ONC centre. It is straight-forward to search 
for these missing OB-stars in OB catalogues.
If they cannot be found then the IMF may be steeper
at the high mass end. Or OB stars form with an initially high binary
fraction so that ejections occur faster and the resulting velocities are
higher, placing them even further away from the ONC centre after 1~(2)~Myr.

Due to the absence of primordial binaries ours is
a conservative result because binaries enhance the ejection
rates. Therefore the influence of 
primordial binaries must be investigated in further
experiments, which will also need to consider gas expulsion from
the embedded cluster \citep{kroupa2001b,vine2003a}. 
\citet{vine2003a} investigated
the evolution of cores of young star clusters and their massive stars
but used a smoothed potential to avoid the difficulties coming
up with the occurrence of close encounters. But these close encounters
are the energy sources for massive star ejections and the reasons
why young stars can be found far away from the cluster centre within
a short period of time. 

\revised{The influence of the cluster potential and especially of
  two-body relaxation with low-mass stars in the cluster shell may
  also be investigated in further simulations. Of what kind this
  influence is is still somewhat unclear.  Stronger constraining forces
  by the cluster potential and energy-loss by two-body relaxation may
  return some of the OB-stars evaporated with low velocities
  from the core. This may stabilize the core on the one hand, but may
  also lead to a faster decay by shifting the velocity spectrum of the
  ejected stars to higher velocities because the stellar density at
  the centre would be higher. This would increase the probability of close
  encounters and high velocity ejections. If relaxation indeed stabilises
  the core then more OB-stars are expected to remain in the ONC
  worsening the OB-stars discrepancy. Ours is therefore a conservative result.}

As a final note, Kroupa et al. (2001) have presented star-cluster-formation
calculations that reproduce the ONC at an age of 1~Myr {\it and} the
Pleiades at an age of 100~Myr. These models, however, are about 1.8-times as
massive as the ONC mass used here ($2200\,M_\odot$) implying that if the IMF
was canonical then the ONC may have had $1.8\times 40 = 72$ stars more
massive than $5\,M_\odot$. This would pose an increased challenge, because
as is evident from Fig. \ref{decay}, 40~OB stars already lead 
to an acceptable match
with the data, so
72~would increase the probability of finding a trapezium-configuration at
an age of 1~Myr, but would lead to too many OB stars within the cluster (as
can be deduced from the lower-panel in Fig. \ref{decay}).

Clearly, such
models with a high initial multiplicity fraction need to be constructed for
further studies of the intricate interrelation of the IMF with
stellar dynamics in young clusters.

\vspace{1cm}
We thank Ian Bonnell for very valuable suggestions.
This work was mainly supported by the GRK-787 Bochum-Bonn 
\emph{Galaxy groups as laboratories for baryonic and dark matter}. 
Jan Pflamm-Altenburg also thanks Gerd Weigelt and the MPIfR 
for financial support at the beginning of this work and especially 
Ralf J\"urgen Dettmar, spokesman of the GRK-787, for important support. 
\bibliographystyle{mn2e}
\bibliography{OB-star,star-formation,ONC,n-body,math,imf}

\appendix
\section{A practical numerical formulation of the IMF}
\subsection{The general IMF}
\label{imf}
When doing research using the IMF a
multi-power-law,
\begin{equation}
\xi(m) = k\!
\left\{\!\!\!\!
    \begin{array}{l@{\;,}l}
      \left(\frac{m}{m_{\mathrm{H}}}\right)^{-\alpha_0}&
      m_{\mathrm{low}}\le m \le m_{\mathrm{H}}\\
      \left(\frac{m}{m_{\mathrm{H}}}\right)^{-\alpha_1}&
      m_{\mathrm{H}}\le m \le m_0\\
      \left(\frac{m_0}{m_\mathrm{H}}\right)^{-\alpha_1}
      \left(\frac{m}{m_0}\right)^{-\alpha_2}&
      m_0\le m \le m_1\\
      \left(\frac{m_0}{m_\mathrm{H}}\right)^{-\alpha_1}
      \left(\frac{m_1}{m_0}\right)^{-\alpha_2}
      \left(\frac{m}{m_1}\right)^{-\alpha_3}&
      m_1\le m \le m_\mathrm{max}\\
    \end{array},
\right.
\end{equation}
with exponents for its canonical form,
\begin{equation}
\begin{array}{l@{\;\;\;,\;\;\;}l}
\alpha_0 = +0.30&0.01 \le m/M_\odot \le 0.08,\\
\alpha_1 = +1.30&0.08 \le m/M_\odot \le 0.50,\\
\alpha_2 = +2.30&0.50 \le m/M_\odot \le 1.00,\\
\alpha_3 = +2.35&1.00 \le m/M_\odot \le +\infty,\\
\end{array}
\end{equation}
\citep{kroupa1993a,reid2002a,kroupa2001a,weidner2004a}
requires many if-statements
in code implementations. Note  however that the canonical IMF
parametrisation 
constitutes a two-part power-law IMF in the stellar
regime: $\alpha_1 = 1.3$ for m $\le$ 0.5 M$_\odot$ and $\alpha_{2,3} = 2.3$
for m $\ge$ 0.5 M$_\odot$.
Here we present a general formulation for
an IMF that avoids these difficulties with IF-statements 
and allows any number of 
mass segments and types of interpolating functions. 
Complicated code constructions containing
if-statements are replaced by two easy loops.
The described method is realised by some very handy 
functions implemented in a shared C-library, and is available on 
request or at the AIfA homepage\footnote{\tt http://www.astro.uni-bonn.de}.
The formulation described below can be
applied to any arbitrary distribution functions for any purpose.

Due to historical reasons multi-power-law IMFs start indexing intervals
and slopes at zero instead of one. For simplicity we here index $n$
intervals from 1 up to $n$. 

Consider an arbitrary IMF
with $n$ intervals
fixed by the mass array $[m_{\mathrm{0}},\ldots,m_{\mathrm{n}}]$ and the 
array of functions 
$f_1$,\ldots,$f_\mathrm{n}$.
So on the i-th interval $[m_{\mathrm{i-1}},m_{\mathrm{i}}]$ the IMF is described 
by the function $f_\mathrm{i}$.
For the case of a multi-power law the functions may be chosen to be
\begin{equation}
f_{\mathrm{i}}(m)= m^{-\alpha_{i}}\;.
\end{equation}
To make this power-law description correspond to the multi-power-law
given above the IMF-slope indices above need to be shifted by one.

The following general IMF-description does not require power-laws for
the functions $f_\mathrm{i}$, but also any kind of function is allowed.
This includes log-normal IMFs \citep{miller1979a,chabrier2003a}.
 
With the two $\Theta$-mappings
\begin{equation}
\Theta_{[\phantom{i}]}(x)=\left\{
\begin{array}{cc}
  1&x\ge0\\
  0&x<0
\end{array}\right.
\end{equation}
and
\begin{equation}
\Theta_{]\phantom{i}[}(x)=\left\{
\begin{array}{cc}
  1&x>0\\
  0&x\le0
\end{array}\right.\;,
\end{equation}
the function
\begin{equation}
\Gamma_{[\mathrm{i}]}(m) = \Theta_{[\phantom{i}]}(m-m_{\mathrm{i}-1})
\Theta_{[\phantom{i}]}(m_\mathrm{i}-m)
\end{equation}
can be defined. 
It is unity on the interval $[m_{\mathrm{i}-1},m_{\mathrm{i}}]$
and zero otherwise.
The complete IMF can be conveniently formulated by
\begin{equation}
\xi(m)=
k\;
\prod_{\mathrm{j}=1}^{\mathrm{n}-1}\;\Delta(m-m_\mathrm{j})\;
\sum_{\mathrm{i}=1}^{\mathrm{n}}\;\Gamma_{[\mathrm{i}]}(m)\;
\Psi_{\mathrm{i}}\;f_{\mathrm{i}}(m)\;,
\end{equation}
where $k$ is a normalisation constant and the array
($\Psi_{1}$,\ldots,$\Psi_{\mathrm{n}}$) is to ensure
continuity at the interval boundaries. They  are defined recursively by
\begin{equation}
\Psi_{1}=1
\;\;\;,\;\;\;
\Psi_{\mathrm{i}} = \Psi_\mathrm{i-1}\;
\frac{f_\mathrm{i-1}(m_{\mathrm{i}-1})}
{f_{\mathrm{i}}(m_{\mathrm{i}-1})}\;.
\end{equation}
For a given mass $m$ the $\Gamma_{[\mathrm{i}]}$ makes all
summands zero except the one in which $m$ lies.
Only on the inner interval-boundaries do both adjoined intervals
give the same contribution to the total value.
The product over 
\begin{equation}
\Delta(x)=\left\{
\begin{array}{cc}
0.5& x=0\\ 
1& x\not=0
\end{array}
\right.
\end{equation}
halves the value due to this double counting at the interval-boundaries.
In the case of $n$ equals one (one single power law), 
the empty product has, by convention, 
the value of unity.

An arbitrary integral over the IMF is evaluated by
\begin{equation}
\int_{a}^{b}\;\xi(m)\;\mathrm{d}m=
\int_{m_{0}}^{b}\;\xi(m)\;\mathrm{d}m-
\int_{m_0}^{a}\;\xi(m)\;\mathrm{d}m\;,
\end{equation}
where the primitive of the IMF is given by
\[
\int_{m_{0}}^{a}\;\xi(m)\;\mathrm{d}m=k
\sum_{\mathrm{i}=1}^{\mathrm{n}}\;\Theta_{]\phantom{i}[}(a-m_{\mathrm{i}})
\;\Psi_{\mathrm{i}}\;\int_{m_{\mathrm{i}-1}}^{m_{\mathrm{i}}}\;
f_{\mathrm{i}}(m)
\;\mathrm{d}m
\]
\begin{equation}
\rule{2.0cm}{0pt}+k\sum_{\mathrm{i}=1}^{\mathrm{n}}\;\Gamma_{[\mathrm{i}]}(a)
\;\Psi_{\mathrm{i}}\;\int_{m_{\mathrm{i}-1}}^{a}\;f_{\mathrm{i}}(m)
\;\mathrm{d}m\;.
\end{equation}

The expressions for the mass content, i.e. $m\;\xi(m)$,
and its primitive are easily obtained by multiplying the above
expressions in the integrals by $m$. 
\subsection{The individual cluster IMF}
\subsubsection{Normalising the IMF}
The IMF denotes the number of stars per mass interval. Therefore
the normalisation depends on the cluster mass.
Here we follow the normalisation strategy by 
\citet{weidner2004a}. This method requires two further
masses. $m_{\mathrm{max}*}$ is the maximum physically possible
stellar mass and $m_\mathrm{max}$ is the expected maximum stellar mass
in a given cluster of mass $M_\mathrm{cl}$. 
With $\xi = k\;\xi_k(m)$ two equations defining $m_{\mathrm{max}*}$ and
$m_\mathrm{max}$ result:
\begin{equation}
M_{\mathrm{cl}} = k\;\int_{m_0}^{m_{\mathrm{max}}}\;m\;\xi_{k}(m)\;\mathrm{d}m,
\end{equation}
\begin{equation}
1=k\;\int_{m_{\mathrm{max}}}^{m_{\mathrm{max}*}}\;\xi_{k}(m)\;\mathrm{d}m.
\end{equation}
To solve these two equations for $k$ and $m_{\mathrm{max}}$ they  can be
divided by each other leading to an expression for the 
cluster mass as a function
of $m_{\mathrm{max}}$
\begin{equation}
M_\mathrm{cl}=\int_{m_0}^{m_{\mathrm{max}}}\;m\;\xi_{k}(m)\;\mathrm{d}m/
\int_{m_{\mathrm{max}}}^{m_{\mathrm{max}*}}\;\xi_{k}(m)\;\mathrm{d}m.
\end{equation}
As a function of $m_{\mathrm{max}}$ it is continuous and 
strictly monotonically increasing and its image is $\mathcal{R}_{\geq 0}$.
Therefore the existence of a solution $m_\mathrm{max}$ is concluded by
a connectivity argument and the uniqueness follows from the strict monotony.
This solution can be gained using any equation-solving method.

Above $m_\mathrm{max}$ no stars can be found and the IMF in a star cluster
can be expressed by
\begin{equation}
\xi_\mathrm{cl}(m) = \Theta_{[\phantom{i}]}(m_\mathrm{max}-m)\;\xi(m)\;.
\end{equation}
\subsubsection{Dicing stars -- the mass-generating function}
When doing research on the IMF using
Monte Carlo simulations or in setting-up star clusters for
$N$-body simulations a finite set of random
masses distributed according to the IMF have to be diced. 
A random number $X$ is drawn 
from a constant distribution and then transformed into a mass
$m$. The mass segments transformed into the $X$-space are fixed by the 
array 
$\lambda_{0}$,\ldots,$\lambda_{n}$
defined by
\begin{equation}
\lambda_{i}=\int_{m_0}^{m_{\mathrm{i}}}\xi_\mathrm{cl}(m)\;\mathrm{d}m.
\end{equation}
If $P(X)$ denotes a constant distribution between 0 and 
$\lambda_\mathrm{n}$, both functions are related by 
\begin{equation}
\int_{m_0}^{m(X)}\xi_\mathrm{cl}(m^\prime)\;\mathrm{d}m^\prime = 
\int_0^XP(X^\prime)\;\mathrm{d}X^\prime = X
\end{equation}
for a uniform distribution $P(X)$.
The solution of this equation for $m$ is given by 
\[
m=\sum_{\mathrm{i}=1}^{\mathrm{n}}
{_\lambda}\Gamma_{[\mathrm{i}]}F_{\mathrm{i}}^{-1}\left(
\frac{X-\lambda_{i-1}}{k\;\Psi_{\mathrm{i}}}+F_{\mathrm{i}}(m_{\mathrm{i}-1})
\right)
\]
\begin{equation}
\rule{0.5cm}{0pt}\cdot
\prod_{\mathrm{j}=1}^{\mathrm{n}-1}\;\Delta(X-\lambda_{\mathrm{i}})
\end{equation}
where $F_{\mathrm{i}}$ is the primitive of 
$f_{\mathrm{i}}$, $F^{-1}_{\mathrm{i}}$ is
the primitive's inverse mapping and $_\lambda\Gamma_{\mathrm{i}}$
are mappings which are unity between $\lambda_{\mathrm{i}-1}$
and $\lambda_{\mathrm{i}}$ and zero otherwise.
\section{Finding the number of expected OB-stars in the ONC} 
\label{finding-N-OB}
The observed total stellar mass of the ONC may be less
than the initial one if OB-stars have been ejected.
The total stellar mass $M_\mathrm{cl}$ is related to
the IMF by
\begin{equation}
M_\mathrm{cl} = M_{<5}+\int_5^{m_\mathrm{max}}m\;\xi(m)\;\mathrm{d}m,
\end{equation}
where $M_{<5}$ is the observed mass in stars less massive 
than 5 $M_\odot$. For a total mass for the ONC of 
1800 - 3300 M$_\odot$ the expected maximum mass
$m_\mathrm{max}$ lies in the range 50 - 63 M$_\odot$
\citep{weidner2006a}.
If 5 $M_\odot < m_\mathrm{max}$, which is the case, the IMF 
can be normalised directly by
\begin{equation}
M_{<5}=k\;\int_{m_0}^{5}m\;\xi_k(m)\;\mathrm{d}m,
\end{equation}
where $m_0 = 0.01\;\mathrm{M}_\odot$ is the opacity-limited
minimum fragmentation mass. 
The maximum stellar mass is then determined in the second step
solving 
\begin{equation}
1=\int_{m_{\mathrm{max}}}^{m_{\mathrm{max}*}}\;\xi(m)\;\mathrm{d}m,
\end{equation}
which means there exists one most massive star in the cluster
\citep{weidner2004a}.
The expected number of OB stars is then given by
\begin{equation}
N_{\mathrm{OB}} = \int_5^{m_\mathrm{max}}\xi_{\mathrm{cl}}\;\mathrm{d}m.
\end{equation}
\end{document}